\documentclass[showpacs,preprintnumbers,amssymb]{revtex4}
\usepackage{epsfig}

\usepackage{threeparttable}

\def\beq{\begin{eqnarray}}
\def\eeq{\end{eqnarray}}
\def\nnb{\nonumber}

\newcommand{\be}{\begin{equation}}
\newcommand{\ee}{\end{equation}}
\newcommand{\bea}{\begin{eqnarray}}
\newcommand{\eea}{\end{eqnarray}}
\newcommand{\ba}{\begin{array}}
\newcommand{\ea}{\end{array}}
\begin{document}
\title{Lepton Flavor Violating $\tau^- \to \mu^- PP$ Decays
in the Two Higgs Doublet Model III}
\author{Wenjun Li}
\email{liwj24@163.com}
 \affiliation{ Department of Physics, Henan
Normal University, XinXiang, Henan, 453007, P.R.China \nnb \\ Kavli
Institute for Theoretical Physics China, CAS, Beijing 100190, China}
\author{ YingYing Fan, Gongwei Liu}
 \affiliation{ Department of Physics,
Henan Normal University, XinXiang, Henan, 453007, P.R.China}

\begin{abstract}
In this paper, the lepton flavor violating $\tau^- \to \mu^-PP
(PP=K^+K^-,K^0\bar{K}^0,\pi^+\pi^-,\pi^0\pi^0)$ decays are studied
in the framework of the two Higgs doublet model(2HDM) III. We
calculate these decays branching ratios and get the bounds of model
parameter $|\lambda_{\tau\mu}|$ from the experimental upper limits.
Our results show that, the neutral Higgs bosons have tree-level
contributions to these decays. Among these decays, the $\tau^- \to
\mu^- K^+K^-$ decay is most sensitive to $|\lambda_{\tau\mu}|$. In
the existing parameters space, these decays could reach the measure
capability of B factory. These processes can provide some valuable
information to future research and furthermore present the reliable
evidence to test the 2HDM III model.

\end{abstract}


\maketitle

\noindent
\section{\bf Introduction}
    Flavor physics have made rapid development in these decades.
In addition to B physics, $\tau$ physics, including determination of
$\alpha_s$ from the inclusive hadronic width, charged-current
universality tests, and lepton flavor violation decays etc., also
belongs these days to one branch of flavor physics. Among these
topics, lepton flavor violation(LFV) decays raises in importance
after the discovery of neutrino flavor oscillations and related
non-zero neutrino masses\cite{superK}. In the SM these processes are
forbidden or suppressed strongly, therefore, LFV decays could be a
sharp tool to seek for some new scenarios with new LFV source and/or
new particles.
The theoretical investigations of $\tau \to 3l,\tau \to \mu \gamma
,\tau \to lP(V^0)$ have sprung up in different
contexts\cite{t3l,tga,tlm,li}. With only two meson in final states,
$\tau^- \to \mu^-PP$ decays are so clean as to provide some
information of QCD. And the experimental upper limit of $\tau \to
K^+K^-(K^0\bar{K}^0,\pi^+\pi^-,\pi^0\pi^0)$ are ~\cite{belle&data}:
\bea &&{\cal B}(\tau^- \to \mu^- K^+K^-)<6.8 \times
10^{-8},\,\,\,\,90\%CL
\nnb \\
&&{\cal B}(\tau^- \to \mu^- \pi^+\pi^-)<3.3 \times
10^{-8},\,\,\,\,90\%CL \nnb
\\&&{\cal B}(\tau^- \to \mu^- K^0_sK^0_s)<3.4 \times
10^{-6}, \,\,\,\,90\%CL
\\&&{\cal B}(\tau^- \to \mu^- \pi^0\pi^0)<1.4 \times
10^{-5}, \,\,\,\,90\%CL
 \eea where the former two values have improved the previous upper
 bounds by almost an order of magnitude.

There are also lots of theoretical researches on $\tau \to l PP$
decays in many possible extensions of the
SM\cite{Ilakovac&chen,Herrero,Arganda,Yue}. For $\tau \to l PP$
decays with hadrons in final states, their amplitudes could be
separated into leptonic vertexes and hadronic parts. One approach to
handling the latter is usually parameterized as hadrons mass and
their decay constants, which could be determined by experimental
values. 
some authors have made analysis of these processes from views of
vector meson dominance, chiral symmetry breaking, Breit-Wigner
propagators, etc.\cite{Ilakovac&chen}. For the hadronisation of
final state in $\tau$ decays, its scale is at the order of $~1GeV$
which lies in the non-perturbative region. Hence, we need consider
the non-perturbative methods one of which is the Chiral Perturbative
Theory($\chi PT$)\cite{Pt}. Different from one pseudoscalar meson in
final state, the resonances are participated in the processes of
$\tau^- \to \mu^- PP$. Stemmed from $\chi PT$, the Resonance Chiral
Theory($R\chi T$) has developed\cite{Rt}. Using $R\chi T$, E.
Arganda $\textit{et al.}$ have investigated these processes in two
constrained MSSM-seesaw scenarios\cite{Arganda}. M.Herrero {\it et
al.} also have make an discussion on the sensitivity of LFV $tau$
decays the Higgs sector of SUSY-seesaw models\cite{Herrero}. The new
particles effects to $\tau^- \to \mu^- PP$ decays  in the TC model
and the LHT model are calculated by Yue chongxing's group\cite{Yue}.

In 2HDM model III, it naturally introduces flavor-changing neutral
currents(FCNCs) at tree level. In order to satisfy the current
experiment constrains, the tree-level FCNCs are suppressed in
low-energy experiments for the first two generation fermions. While
processes concerning with the third generation fermions would be
larger. These FCNCs with neutral Higgs bosons mediated may produce
sizable effects to the $\tau - \mu$ transition. The authors in
\cite{Matsuzaki} have discussed the $\tau \to 3\mu$ decay and the
Higgs sector's contributions. $\tau \to \mu P(V^0)$ decays have been
studied under this scenario in our previous work\cite{li} where the
hadronisation in final state is merely expressed in terms of the
meson  decay constants and meson masses. The $\tau \to \mu P$ decays
could yield one pseudoscalar meson from the vacuum state through the
scalar and pseudoscalar currents. Hence, this type decay could occur
at the tree level through the neutral Higgs bosons exchange. In this
paper, we extend our discussion to the case of two pseudoscalar
mesons in the hadronic final state and deal with these four decays
by $R\chi T$. Our results suggest that, the neutral Higgs bosons
contribute at the tree level in 2HDM model III. The model parameter
$\lambda_{\tau\mu}$ is restrained at $O(10\sim 10^{3})$ and the
decay branching ratios could as large as the current upper limits.

The paper is organized as follows: In section II, we make a brief
introduction of the theoretical framework for the two-Higgs-doublet
model III. In section III, we briefly introduce the Resonance Chiral
 Theory. In the next section, we deliberate the calculation of
the decay amplitudes with Resonance
Chiral Theory and our numerical predictions. Our conclusions are listed in
the last section.

\section{\bf The Two-Higgs -Doublet Model III}
As the simplest extension of the SM, the Two-Higgs-Doublet Model has
an additional Higgs doublet. In order to ensure the forbidden FCNCs
at tree level, it requires either the same doublet couple to the
\textit{u}-type and \textit{d}-type quarks(2HDM I) or one scalar
doublet couple to the \textit{u}-type quarks and the other to
\textit{d}-type quarks(2HDM II). While in the 2HDM
III\cite{cheng,2hdm3}, two Higgs doublets could couple to the
\textit{u}-type and \textit{d}-type quarks simultaneously.
Particularly, without an {\it ad hoc} discrete symmetry exerted,
this model permits flavor changing neutral currents occur at the
tree level.

The Yukawa Lagrangian is generally expressed as the following form:

\beq {\cal L}_{Y}= \eta^{U}_{ij} \bar Q_{i,L} \tilde H_1 U_{j,R} +
\eta^D_{ij} \bar Q_{i,L} H_1 D_{j,R} + \xi^{U}_{ij} \bar
Q_{i,L}\tilde H_2 U_{j,R} +\xi^D_{ij}\bar Q_{i,L} H_2 D_{j,R} \,+\,
h.c., \label{lyukmod3}
 \eeq
 where $H_i(i=1,2)$ are the two Higgs doublets.
 $Q_{i,L}$ is the left-handed
 fermion doublet, $U_{j,R}$ and
$D_{j,R}$ are the right-handed singlets, respectively. These
$Q_{i,L}, U_{j,R}$ and $D_{j,R}$ are weak eigenstates, which can be
rotated into mass eigenstates. While $\eta^{U,D}$ and $\xi^{U,D}$
are the non-diagonal matrices of the Yukawa couplings.

We can conveniently choose a suitable basis to denote $H_1$ and
$H_2$ as: \bea
 \label{base}
 H_1=\frac{1}{\sqrt{2}}\left[\left(\ba{c} 0 \\
v+\phi^0_1 \ea\right)+ \left(\ba{c} \sqrt{2}\, G^+\\
i G^0\ea\right)\right], \,\,\,\,\,\,\,\,
 H_2=\frac{1}{\sqrt{2}}
 \left(\ba{c}\sqrt{2}\,H^+\\ \phi^0_2+i A^0\ea\right),
 \eea
where $G^{0,\pm}$ are the Goldstone bosons, $H^{\pm}$ and $A^0$ are
the physical charged-Higgs boson and CP-odd neutral Higgs boson,
respectively. Its virtue is the first doublet $H_1$ corresponds to
the scalar doublet of the SM while the new Higgs fields arise from
the second doublet $H_2$.

The CP-even neutral Higgs boson mass eigenstates $H^0$ and $h^0$ are
linear combinations of $\phi_1^0$ and $\phi^0_2$ in Eq.(\ref{base}),
 \bea \label{masseigen}
H^0 & = & \phi_1^0 \cos\alpha + \phi^0_2\sin\alpha ,\,\,\, h^0  =
-\phi^0_1\sin\alpha + \phi^0_2 \cos\alpha   ,
 \eeq
 where $\alpha$ is the mixing angle.

After diagonalizing the mass matrix of the fermion fields, the
Yukawa Lagrangian becomes\cite{David}
 \bea  L_Y&=&-\overline{U}M_UU-\overline{D}M_DD
 +\frac{i}{\upsilon}\chi^0\left(
 \overline{U}M_U\gamma_5U-\overline{D}M_D\gamma_5D\right)\nnb \\
 &+&\frac{\sqrt{2}}{\upsilon}
\chi^-\overline{D}V^\dagger_{CKM}\left[M_UR-M_DL\right]U
-\frac{\sqrt{2}}{\upsilon}\chi^+
\overline{U}V_{CKM}\left[M_DR-M_UL\right]D\nnb \\
&+&\frac{iA^0}{\sqrt{2}}\left\{\overline{U}\left[\widehat{\xi}^UR-\widehat{\xi}^{U\dag}L\right]U
+\overline{D}\left[\widehat{\xi}^{D\dag}L-\widehat{\xi}^{D}R\right]D\right\}\nnb\\
 &-&\frac{H^0}{\sqrt{2}}\overline{U}\left\{\frac{\sqrt{2}}{\upsilon} M_U
 \cos\alpha+\left[\widehat{\xi}^UR+\widehat{\xi}^{U\dag}L\right]\sin\alpha\right\}U
 -\frac{H^0}{\sqrt{2}}\overline{D}\left\{\frac{\sqrt{2}}{\upsilon} M_D
 \cos\alpha+\left[\widehat{\xi}^DR+\widehat{\xi}^{D\dag}L\right]\sin\alpha\right\}D\nnb\\
 &-&\frac{h^0}{\sqrt{2}}\overline{U}\left\{-\frac{\sqrt{2}}{\upsilon} M_U
 \sin\alpha+\left[\widehat{\xi}^UR+\widehat{\xi}^{U\dag}L\right]\cos\alpha\right\}U
-\frac{h^0}{\sqrt{2}}\overline{D}\left\{\frac{\sqrt{2}}{\upsilon}
M_D
 \sin\alpha+\left[\widehat{\xi}^DR+\widehat{\xi}^{D\dag}L\right]\cos\alpha\right\}D\nnb\\
 &-&H^+\overline{U}\left[V_{CKM}\widehat{\xi}^DR-\widehat{\xi}^{U\dag}V_{CKM}L\right]D
 -H^-\overline{D}\left[\widehat{\xi}^{D\dag}V^\dagger_{CKM}L-V^\dagger_{CKM}\widehat{\xi}^UR\right]U
 \label{lyukmass}
  \eea
where U and D now are the fermion mass eigenstates and
 \bea
\hat\eta^{U,D}&=&(V_L^{U,D})^{-1}\cdot \eta^{U,D} \cdot
V_R^{U,D}=\frac{\sqrt{2}}{v}M^{U,D}(M^{U,D}_{ij}=\delta_{ij}m_j^{U,D}),
\label{diag}\\
\hat\xi^{U,D}&=&(V_L^{U,D})^{-1}\cdot \xi^{U,D} \cdot V_R^{U,D}
\label{neutral},
 \eea
 where $V_{L,R}^{U,D}$ are
the rotation matrices acting on up and down-type quarks, with left
and right chiralities respectively. Thus
$V_{CKM}=(V_L^U)^{\dag}V_L^D$ is the usual Cabibbo-Kobayashi-Maskawa
(CKM) matrix. In general, the matrices $\hat\eta^{U,D}$ of
Eq.(\ref{diag}) are diagonal, while the matrices $\hat\xi^{U,D}$ are
 non-diagonal which could induce scalar-mediated FCNC. Seen
from Eq.({\ref{lyukmass}}), the coupling of neutral Higgs bosons to
the fermions could generate FCNC parts.
  For the arbitrariness of definition for $\xi^{U,D}_{ij}$ couplings,
we can adopt the rotated couplings expressed $\xi^{U,D}$ in stead of
$\hat{\xi}^{U,D}$ hereafter.

In this work, we use the Cheng-Sher ansatz\cite{cheng} \be
\xi^{U,D}_{ij}=\lambda_{ij} \,\frac{\sqrt{2}\sqrt{m_i m_j}}{v}\label{ans}
\ee
 which ensures that the FCNCs within the first two
generations are naturally suppressed by small fermions masses. This
ansatz suggests that LFV couplings involving the electron are
 suppressed, while LFV transitions involving muon and tau
are much less suppressed and may lead to some loop effects which are
promising to be tested by the future B factory experiments. In
Eq.(\ref{ans}), the parameter $\lambda_{ij}$ is complex
 and $i,j$ are the generation indexes. In this study, we shall discuss the phenomenological
applications of the type III 2HDM.

\section{\bf The Resonance Chiral Theory}
For the intermediate and low energy parts of hadronic spectrum, they
locate at the non-perturbative region where ordinary perturbative
QCD methods does not work. Hence, many non-perturbative approaches
have been developed,such as $\chi PT$\cite{Pt}, QCD sum
rules\cite{Sum}, lattice gauge theory\cite{lat} and so on. The
large-$N_C$ expansion of $SU(N_C)$ QCD\cite{LNC} is a suitable idea.

$\chi PT$ is a appropriate method with $1/N_C$ expansion which is
very successful in the energy reign of $\simeq 1GeV$. It could deal
with $\tau^- \to \mu^- P$ decays which will be discussed in our
later paper. While for $\tau^- \to \mu^- PP$ processes, the
resonances paly a dynamical role so that we should by the aid of
$R\chi T$\cite{Rt}. Motivated by partially $\chi PT$ and large $N_c$
QCD, $R\chi T$ could describe the immediate energy region. Its
advantages are not only to realize the non-linear of spontaneous
chiral symmetry breaking, but also to study meson
 structure without imposing any structure in advance.
  Some application on $R\chi T$ have been elaborated in
\cite{Herrero,Arganda,trnka}. For energies $1\sim 2GeV$, we restrict
it to only the lightest resonance in each channel. The detailed
deliberation can be found in Ref.\cite{Arganda}.

The symmetry of QCD breaks spontaneously from $SU(3)R \bigotimes
SU(3)_L$ to $SU(3)_V$ and produces eight Goldstone bosons in the
spectrum. We regard these Goldstone bosons as the lightest hadrons.
The $\chi PT$ Lagrangian can be constructed
 from pseudoscalar fields and external
source $v^i_\mu(x), a^i_\mu(x), s_i(x), p^i(x)$:
 \bea
 {\cal L}&=&{\cal L}_0+{\cal L}_1,\nnb \\
{\cal L}_0&=&-\frac{1}{2g^2}Tr G_{\mu\nu}G^{\mu\nu}+\bar{q}i
\gamma^\mu(\partial_\mu-iG_\mu) q,\,\,\,\,
{\cal L}_1=\bar{q} [\gamma_\mu (v_\mu + \gamma_5 a^\mu)
-(s-ip\gamma_5)] q,\nnb \\
v^\mu&=&v_i^\mu\frac{\lambda_i}{2},\,\,\,
a^\mu=a^\mu_i\frac{\lambda_i}{2},\,\,\,
s=s_i \lambda_i,\,\,\,
p=p_i \lambda_i,\
 \eea
where the ${\cal L}_1$ is the massless QCD Lagrangian. $G_{\mu\nu}$
denotes the gluon fields, $v^\mu, a^\mu, s, p$ are matrices in the flavor fields
and $\lambda_i$s are the Gell-Mann matrices. The QCD generating functional
$\mathcal{Z}_{QCD}[v,a,s,p]$ could be written:
\begin{equation}
e^{i\mathcal{Z}_{QCD}[\upsilon,a,s,p]}=\int[DG_{\mu}][Dq][D\bar{q}]e^{i\int d^{4}x
\mathcal{L}_{QCD}[q,\bar{q},G,\upsilon,a,s,p]}
\end{equation}

We introduce the lightest $U(3)$ nonet of pseudoscalar mesons:
\bea
\phi(x)&=&
\sum^8_{a=0}\frac{\lambda_a}{\sqrt{2}}\varphi_a\nnb \\
&=&\left(
\begin{array}{ccc}
 \frac{1}{\sqrt{2}} \pi^0+\frac{1}{\sqrt{6}}\eta^8+\frac{1}{\sqrt{3}}\eta^0&
  \pi^+ & K^+ \\
\pi^- & -\frac{1}{\sqrt{2}}\pi^0 +\frac{1}{\sqrt{6}}\eta^8+\frac{1}{\sqrt{3}}\eta^0 & K^0 \\
 K^- & \bar{K}^0 & -\frac{2}{\sqrt{6}}\eta^8+\frac{1}{\sqrt{3}}\eta^0
\end{array}
\right) \eea
The unitary $3\times 3$ matrix $u(x)$ can be presented
as: \be u(x)=e ^{i\frac{\phi(x)}{\sqrt{2}F}} \ee

The leading $O(p^2) \chi PT SU(3)_L\times SU(3)_R$ chiral Lagrangian
is \bea {\cal L}^{(2)}_\chi &=&\frac{F^2}{4}\langle u_\mu u^\mu
+\chi_+\rangle \nnb \\
 u_\mu &=&i[u^\dag (\partial_\mu
-ir_\mu)u-u(\partial_\mu -il_\mu)u^\dag], \,\,\,\, \chi_+=
u^\dag\chi u^\dag+ u\chi^\dag u, \,\,\,\, \chi = 2B_0(s + ip). \eea

Interactions with electroweak bosons can
be accommodated through the vector $v_\mu=\frac{(\gamma_\mu+l_\mu)}{2}$
 and axial-vector $a_\mu=\frac{(\gamma_\mu-l_\mu)}{2}$
external fields. The effective coupling constant $F$ is approximately equal the pion decay constant
 of $\pi$. In the chiral limit there has $B_0F^2=-\langle0|\bar{\psi} \psi|0\rangle_0$.
The chiral tensor $\psi$ provides masses to the Goldstone
bosons through the external scalar field $s={\cal M}+\cdots, {\cal M}=\left(
\begin{array}{ccc}
 m_u &
   &  \\
 & m_d &  \\
& & m_s
\end{array}
\right)$.

Indeed in the isospin limit we have:
\bea \chi&=&2B_0{\cal M} + \cdots
=\left(
\begin{array}{ccc}
 m^2_\pi &
   &  \\
 & m^2_\pi &  \\
& & 2m^2_K-m^2_\pi
\end{array}
\right)+\cdots \nnb \\
B_0m_u&=&B_0m_d=\frac{1}{2}m^2_\pi,\,\,\,\,
B_0m_s=m^2_K-\frac{1}{2}m^2_\pi \eea
The mass eigenstates $\eta$ and $\eta'$
are related to the octet $\eta_8$ and singlet $\eta_0$ states
through the rotation matrix:
\bea
\left(
\begin{array}{c}
 \eta  \\
 \eta'
\end{array}
\right)=\left(
\begin{array}{cc}
 \cos\theta& -\sin \theta  \\
 \sin \theta&\cos\theta
\end{array}
\right)\left(
\begin{array}{c}
 \eta_8  \\
 \eta_0
\end{array}
\right)\eea
where $\theta=18^0$.

In the processes of $\tau$ decaying to two pseudo-scalars, vector
and scalar resonances generally play a propelling role. However, due
to the higher masses of scalar resonances, their effects are
ignored\cite{Arganda}. Then we carry out antisymmetric tensor fields
to introduce vector resonances. The nonet of resonance fields
$V_{\mu\nu}$ reads: \bea \left(
\begin{array}{ccc}
 \frac{1}{\sqrt{2}}\rho^0
 +\frac{1}{\sqrt{6}}\omega_8+\frac{1}{\sqrt{3}}\omega_0
 &\rho^+&K^{*+}\\
 \rho^-&-\frac{1}{\sqrt{2}}\rho^0
 +\frac{1}{\sqrt{6}}\omega_8+\frac{1}{\sqrt{3}}\omega_3&K^{*0}\\
 K^{*-}&\bar{K}^{*0}&-\frac{2}{\sqrt{6}}\omega_8+\frac{1}{\sqrt{3}}\omega_0
\end{array}
\right)_{\mu\nu}\eea

Then the $R\chi T$ Lagrangian puts:
\bea
{\cal L}_{R\chi T} &=&{\cal L}^{(2)}_\chi +{\cal L}^V_{(2)},\,\,\,\,
{\cal L}_V ={\cal L}^V_{kin}+{\cal L}^V_{(2)} ,\nnb \\
{\cal L}^V_{kin}&=&\frac{1}{2}\langle\nabla^\lambda V_{\lambda\mu}
\nabla_\nu V^{\nu\mu}\rangle+\frac{M^2_V}{4}\langle
V_{\mu\nu}V^{\mu\nu}\rangle,\,\,\,\, {\cal
L}^V_{(2)}=\frac{F_V}{2\sqrt{2}}\langle
V_{\mu\nu}f^{\mu\nu}_+\rangle +i\frac{G_V}{\sqrt{2}}\langle
V_{\mu\nu} u^\mu u^\nu\rangle \eea where ${\cal L}_V$ is the
resonance Lagrangian. The relevant definitions can be found in
Ref.\cite{Arganda}. The corresponding QCD generating functional is
written as \bea e^{iZ_{QCD}[v,a,s,p]}=\int [Du][DV]e^{i\int d^4x
{\cal L}_{R\chi T}[u,V,v,a,s,p]} \eea
Through making the proper
partial derivatives of the functional action, we can get the
hadronisation of bilinear quark currents: \bea V^{i}_{\mu} &=&
\bar{q}\gamma_{\mu}\frac{\lambda^{i}}{2}q=\frac{\partial
\mathcal{L}_{R\chi T}}
  {\partial \upsilon ^{\mu}_{i} }|_{j=0},\,\,\,\,
  A^{i}_{\mu} =\bar{q}\gamma_{\mu}\gamma_{5}\frac{\lambda^{i}}{2}q=\frac{\partial \mathcal{L}_{R\chi T}}
  {\partial a ^{\mu}_{i} }|_{j=0},\nonumber\\
S^{i}&=& -\bar{q}\lambda^{i}q=\frac{\partial \mathcal{L}_{R\chi T}}
  {\partial s_{i} }|_{j=0}, \,\,\,\, P^{i} = \bar{q}i \gamma_{5}\lambda^{i}q=\frac{\partial \mathcal{L}_{R\chi T}}
  {\partial p^{\mu}_{i} }|_{j=0}\label{current}
\eea The final results from Eq.{(\ref{current})} are \bea
V^{i}_{\mu} &=& \frac{F^{2}}{4}<\lambda
^{i}(uu_{\mu}u^{\dagger}-u^{\dagger}u_{\mu}u)>-\frac{F_{V}}{2\sqrt{2}}
<\lambda^{i}\partial^{\mu}(u^{\dagger}V_{\mu\nu}u+uV_{\mu\nu}u^{\dagger})>\\
A^{i}_{\mu} &=& \frac{F^{2}}{4}<\lambda ^{i}(uu_{\mu}u^{\dagger}+u^{\dagger}u_{\mu}u)> \\
S^{i} &=& \frac{1}{2}B_{0} F^{2}<\lambda ^{i}(u^{\dagger}u^{\dagger}+uu)>\label{pes1} \\
P^{i} &=& \frac{i}{2}B_{0} F^{2}<\lambda
^{i}(u^{\dagger}u^{\dagger}-uu)> \label{pes2}\eea

\section{\bf The discussion for $\tau^- \to \mu^- PP$ decays }
In 2HDM model III, the neutral Higgs bosons mediated tree
 diagrams have contributions to $\tau^- \to \mu^- PP$
processes. The amplitudes could be factorized into leptonic vertex
corrections and hadronic parts described with hadronic matrix
elements, which express as: \bea \langle \mu^-PP |{\cal M} |\tau^-
\rangle &=&\frac{iG_F}{2\sqrt{2}} \cdot m_q\sqrt{m_\tau m_\mu}\cdot
\left\{\biggl[ H^{q*}\cdot \lambda_{\tau\mu} \cdot
(\bar{\mu}\tau)_{S+P}+ H^{q}\cdot \lambda^*_{\tau\mu}\cdot
(\bar{\mu}\tau)_{S-P}\biggl]\cdot<PP|(\bar{q}q)_S|0>
 \right.\nnb
\\&&\left.+\biggl[ N^{q*}\cdot \lambda_{\tau\mu}\cdot
(\bar{\mu}\tau)_{S+P}- N^{q}\cdot \lambda^*_{\tau\mu}\cdot
(\bar{\mu}\tau)_{S-P}\biggl]\cdot<PP|(\bar{q}q)_P|0> \right\}
\label{amp}\eea

where $H^{q},N^q$ are auxiliary functions can be find in Appendix.
From the Eq.(\ref{amp}), we can see that three neutral Higgs bosons
perform roles to two pseudo-scalar mesons through $(\bar{q}q)_{S\pm
P}$ operators. It differs from the case of MSSM models where the
$\gamma$ contributions is the dominate one and only $H^0$ and $h^0$
take effects at the large $\tan \beta$\cite{Arganda}. For the heavy
Higgs bosons, the hadronic final state is not sensitive to
resonances and known little. It should be noted that the
pseudo-scalars currents have contributions to only one pseudo-scalar
 meson in final states. So we have dealt with
 the hadronic matrix
 elements by virtue of Eq.(\ref{pes1}),(\ref{pes2}) and the following
 currents:
 \bea
-\bar{u} u&=&\frac{1}{2}S^3+\frac{1}{2\sqrt{3}}S^8
+\frac{1}{\sqrt{6}}S^0,\nnb \\
-\bar{d}d&=&-\frac{1}{2}S^3+\frac{1}{2\sqrt{3}}S^8
+\frac{1}{\sqrt{6}}S^0,\nnb \\
-\bar{s} s&=&-\frac{1}{\sqrt{3}}S^8 +\frac{1}{\sqrt{6}}S^0 \eea

The obtained amplitudes read as: \vskip-1cm \bea {\cal M}(\tau^- \to
\mu^- PP )&=& \frac{i G_F}{2\sqrt{2}}\cdot \sqrt{m_\tau m_\mu} \cdot
[T(PP) \cdot \lambda_{\tau\mu}\cdot (\bar{\mu}\tau)_{S+P} +T^*(PP)
\cdot \lambda^*_{\tau\mu}\cdot (\bar{\mu}\tau)_{S-P}] \eea where\bea
T(K^+K^-)&=&A\cdot[Re(\lambda_{uu})\cdot m^2_\pi+
 Re(\lambda_{ss})\cdot (2m^2_K
-m^2_\pi)]+Bi\cdot [Im(\lambda_{ss})\cdot (2m^2_K
-m^2_\pi)-Im(\lambda_{uu})\cdot m^2_\pi]\nnb \\&&+F\cdot m^2_K
 \label{amp1}\\
T(K^0\bar{K}^0)&=&A\cdot[Re(\lambda_{dd})\cdot m^2_\pi +Re(\lambda_{ss})\cdot
(2m^2_K -m^2_\pi)]+Bi\cdot[Im(\lambda_{ss})\cdot (2m^2_K -m^2_\pi)+Im(\lambda_{dd})\cdot m^2_\pi
]\nnb \\&&+F\cdot m^2_K \label{amp2}\\
T(\pi^+\pi^-)&=&m^2_\pi\cdot
\left\{A[Re(\lambda_{uu})+Re(\lambda_{dd})]
 +Bi\cdot[ Im(\lambda_{dd})-Im(\lambda_{uu})]+F\right\}
\label{amp3}  \\
T(\pi^0\pi^0)&=&\frac{m^2_\pi}{2\sqrt{2}}\cdot
\left\{A[Re(\lambda_{uu})-Re(\lambda_{dd})]
 -Bi\cdot[ Im(\lambda_{uu}+Im(\lambda_{dd}))]\right\}
\label{amp1}\\
A&=&\frac{\sin^2 \alpha}{m^2_{H^0}}+ \frac{\cos^2
\alpha}{m^2_{h^0}},\,\,\,\,\,\, B=\frac{1}{m^2_{A^0}},\,\,\,\,\,\,
F=2\sin \alpha \cos\alpha (\frac{1}{m^2_{H^0}}-
\frac{1}{m^2_{h^0}})\nnb \eea

 \begin{table}[htb]\label{ql}
 \centering
  \begin{threeparttable}[b]
 \caption{Constraints on the $\lambda_{ij}$ in quark and lepton
sector.}
\begin{tabular}{|c|c|c|c|}
\hline &Bounds and
restrictions& Process and Restriction& References \\
\cline{2-4}
  &$|\lambda_{uu}|,|\lambda_{dd}| \sim O(1)$ &$F^0-\bar{F^0}$ mixing ($F=K,B_d,D$),$R_b,\rho,B\to X_s \gamma$&~\cite{Atwood}\\
  \cline{2-4} &$|\lambda_{tt}|= 0.02,|\lambda_{bb}|= 50$
&$B_d-\bar{B}_d$ mixing ,$b \to s \gamma$& ~\cite{dyb}\\
\cline{2-4}
 Quark sector &$|\lambda_{tt}|<0.5,|\lambda_{bb}|<70,|\lambda_{tt}\lambda_{bb}|\sim 3,|\lambda_{ss}|\in [80,120]$
& $B_d-\bar{B}_d$ mixing ,$b \to s \gamma,\rho_0,R_b,NEDM$&
~\cite{csh}\\
 \cline{2-4} &$|\lambda_{tt}|=|\lambda_{tc}|=0.1,
 |\lambda_{bb}|=|\lambda_{bs}|=50$&$h^0 \to f\bar{f} $& ~\cite{Martin}\\
\cline{2-4}&
$|\lambda_{tt}|=0.3,|\lambda_{bb}|=35,\lambda_{ij}=0$&$B \to
PP,PV,VV$
  & ~\cite{xiao}\\
 \hline
  \cline{2-4}
&$|\lambda_{\tau\tau}|=|\lambda_{\mu\mu}|=5, 50,$
&$B_{d,s}\to l^+l^-$& ~\cite{dyb} \\
\cline{2-4} &$|\lambda_{\mu\mu}|=|\lambda_{\tau\tau}|=
 |\lambda_{\mu\tau}|=|\lambda_{e\mu}|=10$&
 $h^0 \to f\bar{f}$& ~\cite{Martin} \\
 \cline{2-4}
&$\lambda_{\tau\mu}\sim O(10)-O(10^2)$&$(g-2)_\mu,m_{A^0}
\longrightarrow \infty$& ~\cite{Rodolfo03} \\
 \cline{2-4}
 &$\lambda_{\tau\mu} \sim O(10^2)-O(10^3)$&$\tau \to 3\mu,\tau \to \mu \gamma$&~\cite{Cotti}\\
\hline
\end{tabular}
  \begin{tablenotes}
    \item [1] Note that the constraints in\cite{Rodolfo03} and
\cite{Cotti} are denoted by our notation.
   \end{tablenotes}
  \end{threeparttable}
\end{table}

\section{Numerical Results}
    In our calculation, the input
parameters are the Higgs masses, mixing angle $\alpha$,
 $|\lambda_{ij}|$ and their phase angles $\theta_{ij}$.
We using the values of neutral Higgs boson masses in
literature\cite{dyb,csh,Rodolfo03}, where the experimental
constraints of $B-\bar{B}$ mixing, $b\to s \gamma, \rho^0,R_b$
considered. \vskip-1cm \bea  m_{H^0}&=&160GeV,
\,\,\,m_{h^0}=115GeV,\,\,\,
  m_{A^0}=120GeV,\,\,\,\alpha=\pi/4,\,\,\, \theta= \pi/4
  \eea
According the mesons quark constants in final states, the involved
factors of quark sector are $\lambda_{uu},\lambda_{dd}$ and
$\lambda_{ss}$. The bounds of $|\lambda_{\tau\mu}|$ from different
phenomenological considerations\cite{Rodolfo03,Martin,Cotti} are
demonstrated in Tab.I, too. For the first generation FC couplings
are suppressed, the values of $\lambda_{uu}$ and $\lambda_{dd}$ are
less than 1\cite{Atwood}. And the $B_0-\bar{B}_0$ mixing constrains
approximately $\lambda_{ss}$ in $80 \sim 120$\cite{csh}. These
bounds are considered in our calculation. In the following
paragraphs, we will analysis the relations of these decays branching
ratios and relevant parameters . First, we take
$|\lambda_{\tau\mu}|=5$ and study the relations of branching ratio
and other parameters.
\begin{figure}[thbp]
 \includegraphics[scale=0.55]{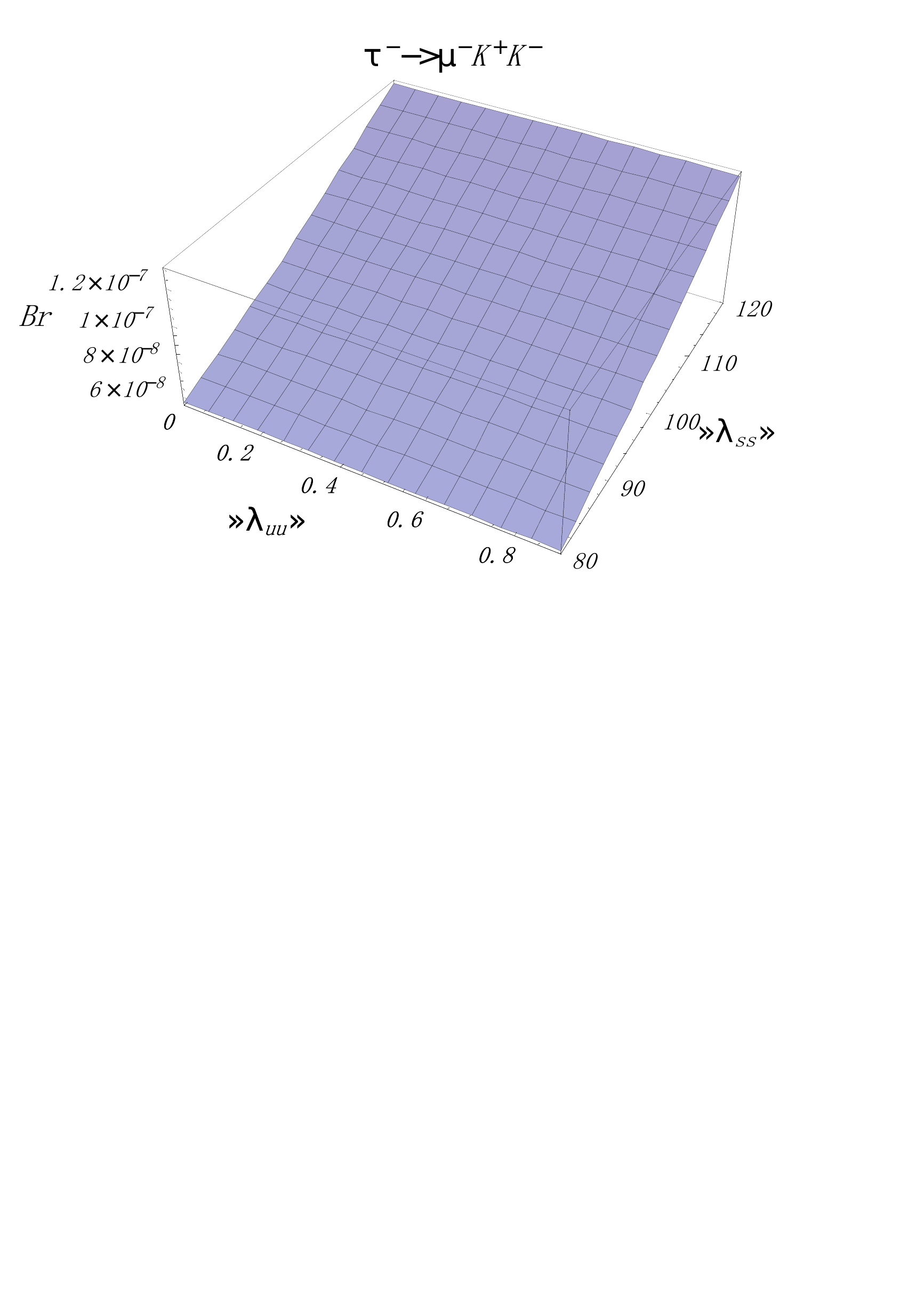}%
\includegraphics[scale=0.55]{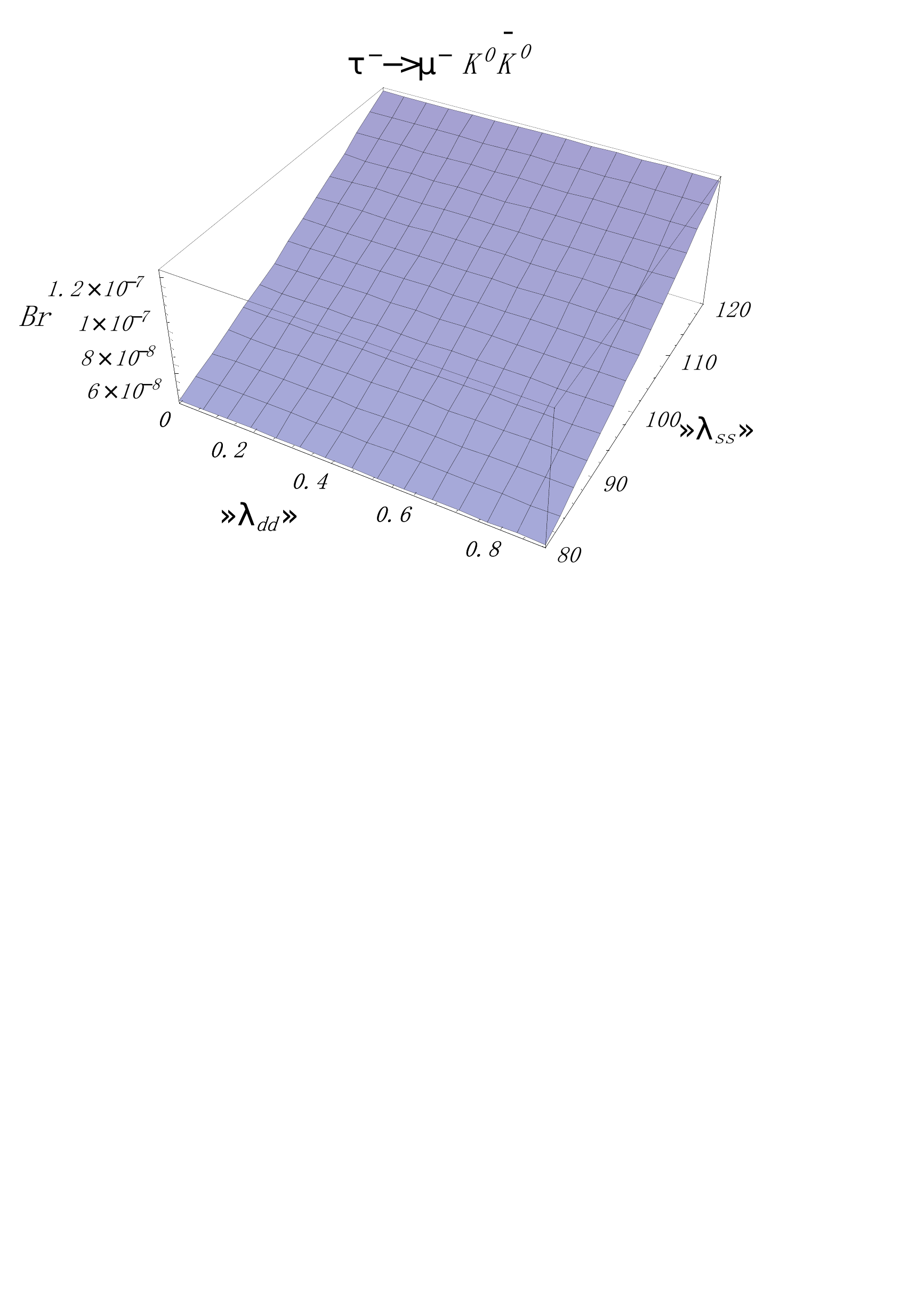}
 \vskip-8cm\caption{Left:$Br$ for $\tau^- \to \mu^- K^+K^-$ versus
 $|\lambda_{uu}|$ and  $|\lambda_{ss}|$;
Right:$Br$ for $\tau^- \to \mu^- K^0\bar{K}^0$ versus
 $|\lambda_{dd}|$ and  $|\lambda_{ss}|$ with $|\lambda_{\tau\mu}|=5$. }
 \end{figure}
 \begin{figure}[thbp]
 \includegraphics[scale=0.55]{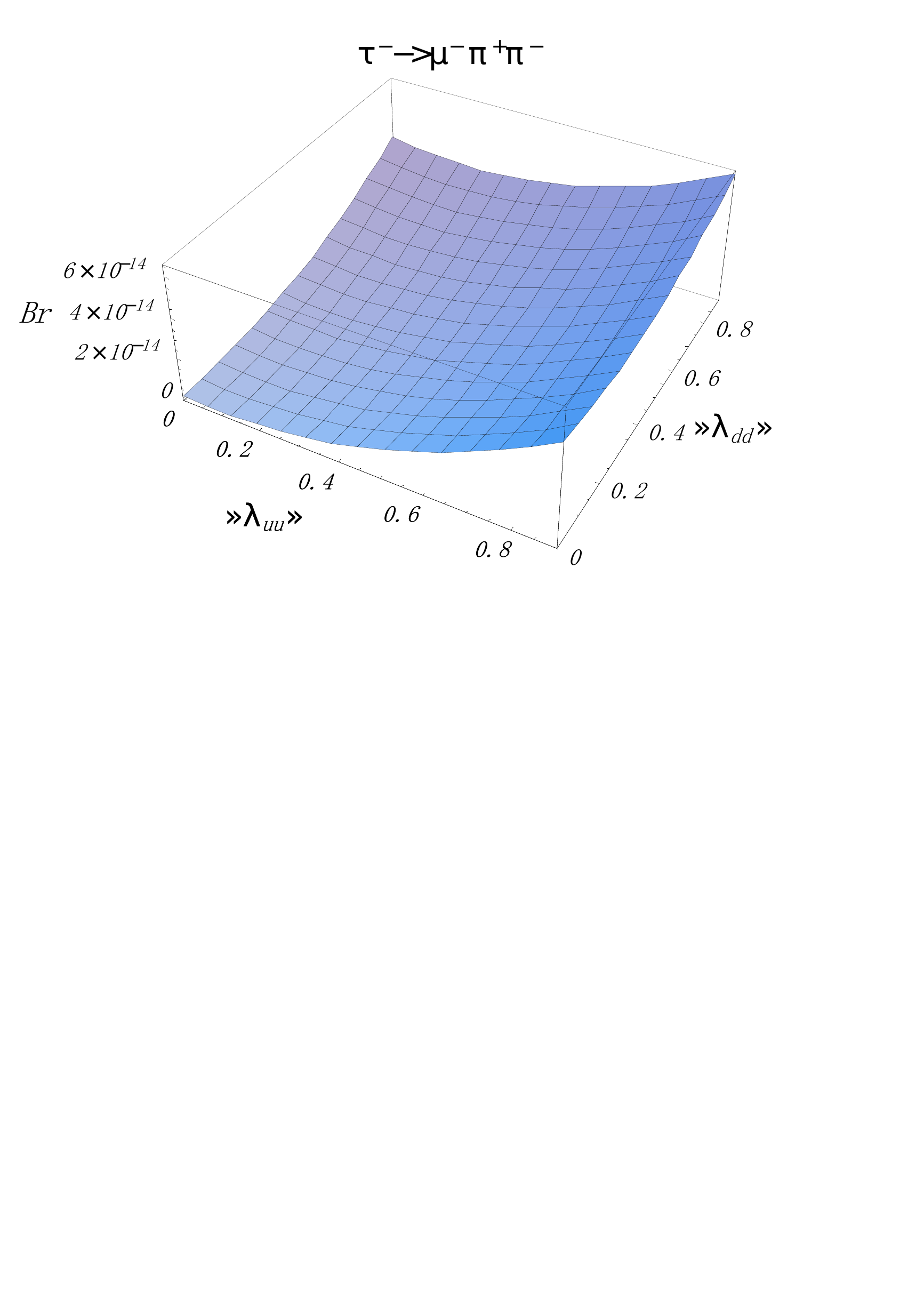}%
\includegraphics[scale=0.55]{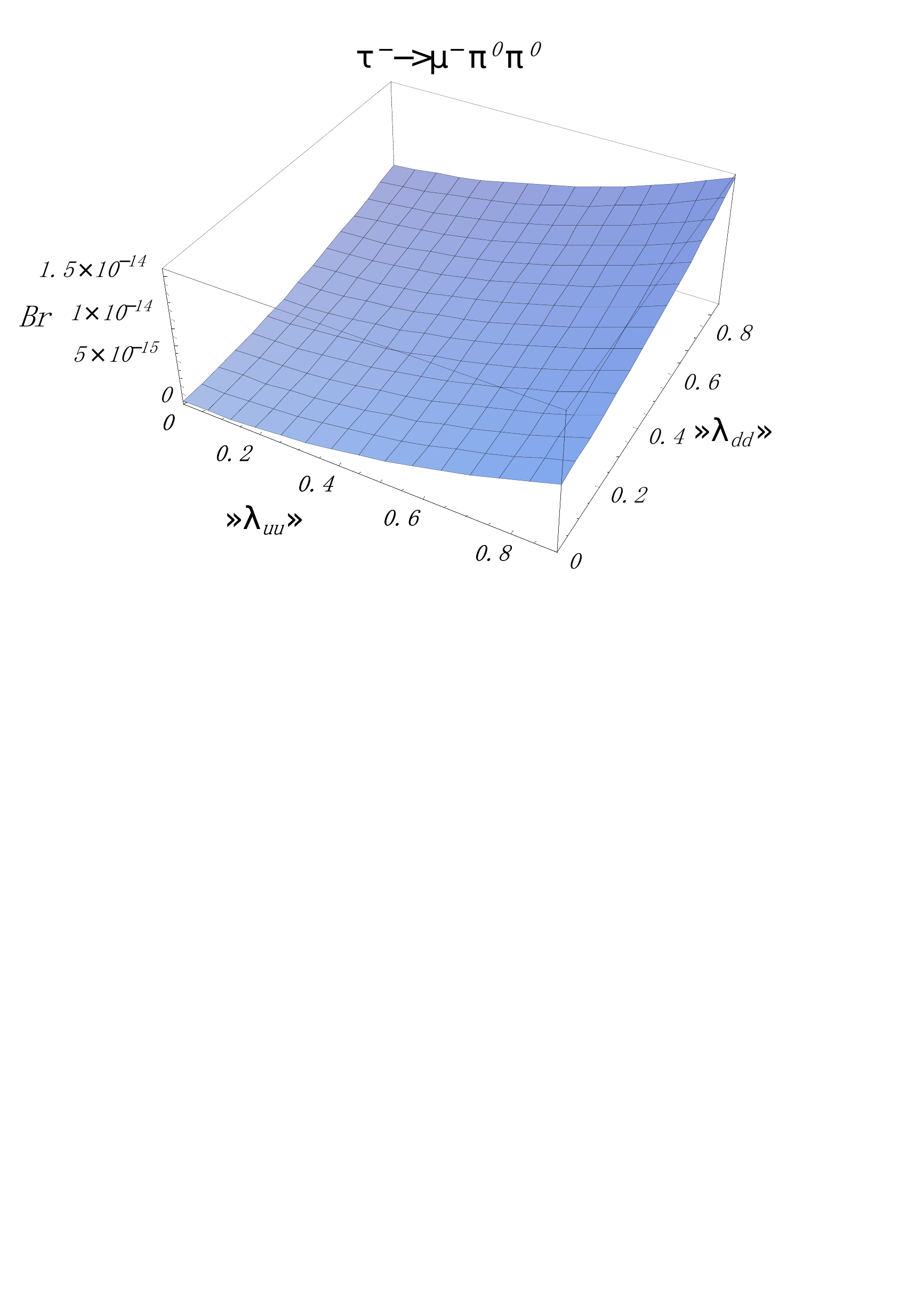}
 \vskip-8cm\caption{Left:$Br$ for $\tau^- \to \mu^- \pi^+\pi^-$ versus
 $|\lambda_{uu}|$ and  $|\lambda_{dd}|$;
Right:$Br$ for $\tau^- \to \mu^- \pi^0\pi^0$ versus
 $|\lambda_{uu}|$ and  $|\lambda_{dd}|$ with $|\lambda_{\tau\mu}|=5$. }
 \end{figure}

\begin{figure}[thbp]
\includegraphics[scale=0.32]{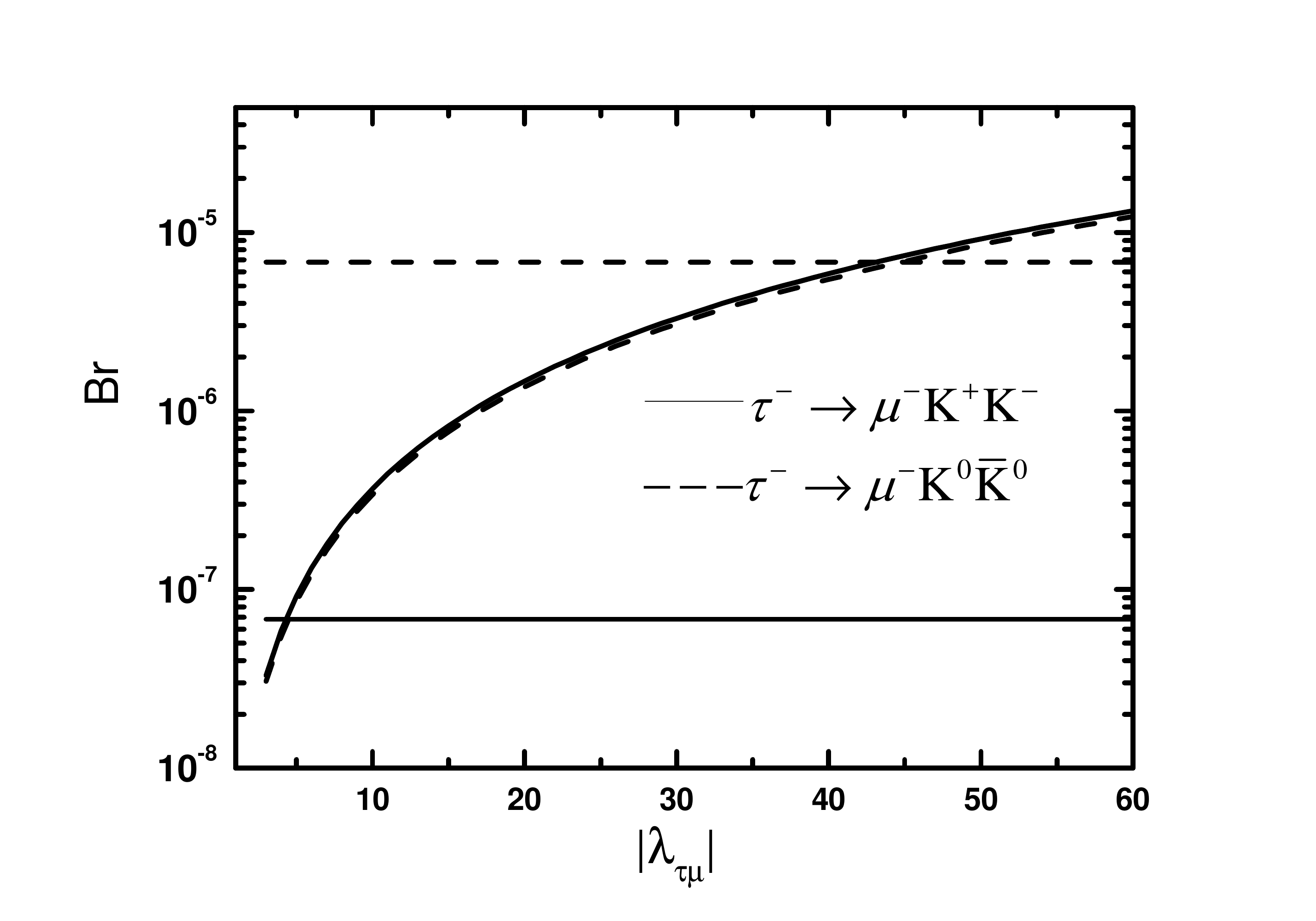}%
\includegraphics[scale=0.32]{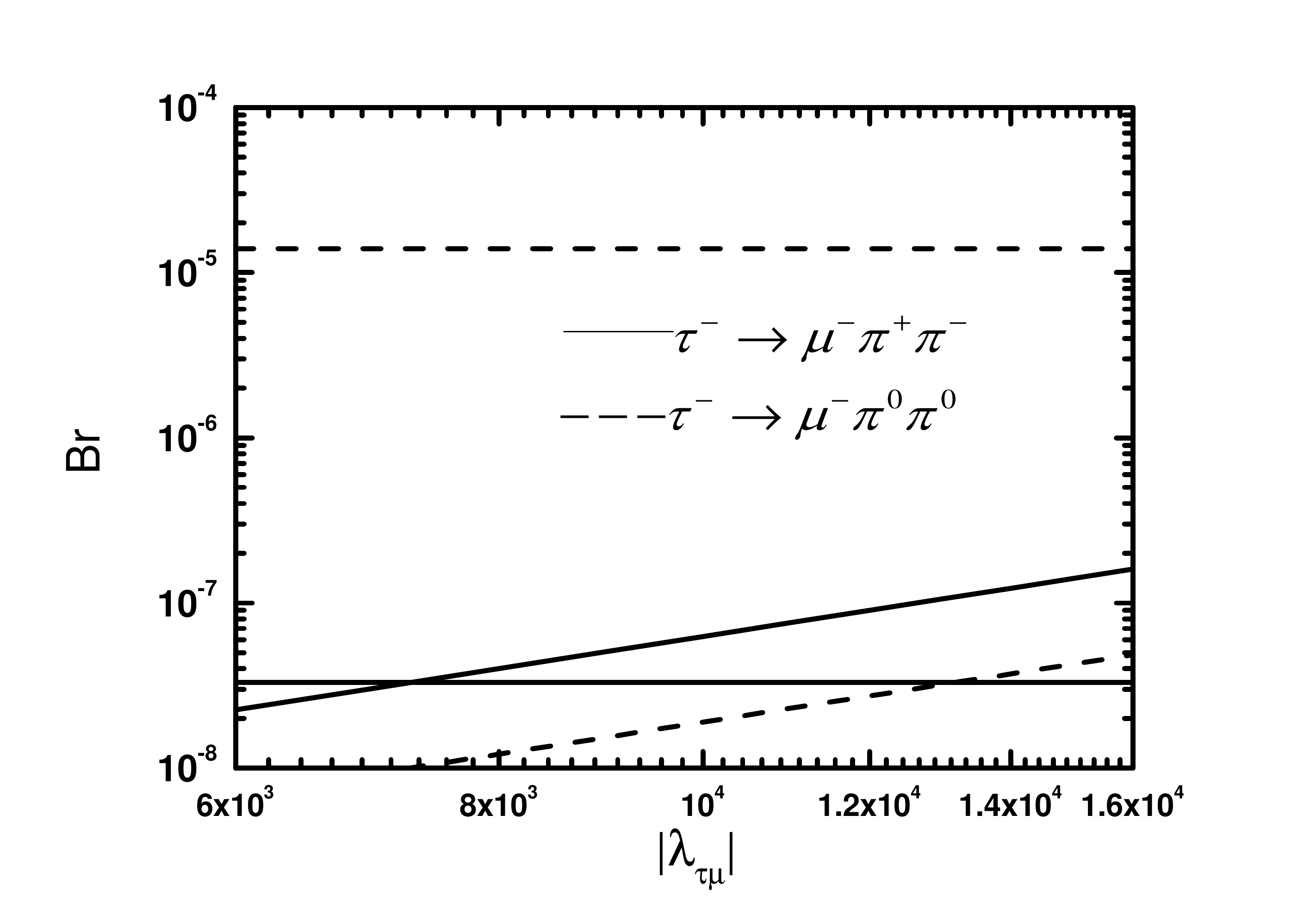}\label{fig4}
\caption{Left: $Br$ for $\tau^- \to \mu^- K\bar{K}$ versus
 $|\lambda_{\tau\mu}|$ for $|\lambda_{uu}|=0.5,|\lambda_{ss}|=100$.
 The solid line stands for $\tau^- \to \mu^- K^+K^-$, the dashing line for
 $\tau^- \to \mu^- K^0\bar{K}^0$,
 the horizon line for current experimental upper limit for $\tau^- \to \mu^- K^+K^-$.
Right:$Br$ for $\tau^- \to \mu^- \pi\pi$ versus
 $|\lambda_{\tau\mu}|$ for $|\lambda_{uu}|=|\lambda_{dd}|=0.5$.
 The solid line stands for $\tau^- \to \mu^- \pi^+\pi^-$, the dashing line for
 $\tau^- \to \mu^- \pi^0\pi^0$,
 the horizon line for current experimental upper limit for
 $\tau^- \to \mu^- \pi^+\pi^-$.}
\end{figure}

  The Fig.1 shows $Br$ for $\tau^- \to \mu^- K  \bar{K}$ decays versus
model parameters, where the left figure is $Br(\tau^- \to \mu^-
K^+K^-)$ versus $|\lambda_{uu}|$ and $|\lambda_{ss}|$, and the right
is $Br(\tau^- \to \mu^- K^0\bar{K}^0)$ versus $|\lambda_{dd}|$ and
$|\lambda_{ss}|$. One can see from Fig.1(a) that $Br(\tau^- \to
\mu^- K^+K^-)$ raises with the increase of $|\lambda_{ss}|$ but does
not vary with $|\lambda_{uu}|$ growing. For $\tau^- \to \mu^-
K^0\bar{K}^0$ decay, the same as that of $\tau^- \to \mu^- K^+K^-$
decay except that $|\lambda_{dd}|$ replaces $|\lambda_{uu}|$. In the
mentioned parameter spaces, both two decays ratios could reach the
order of $10^(-8)$. Comparing Eq.(24) and (25), we could find that
the structures of these decay amplitudes are similar. Hence
$Br(\tau^- \to \mu^- K^0\bar{K}^0)$ also rises with the increase of
 $|\lambda_{ss}|$ but is almost not affected by the modification
 of $|\lambda_{dd}|$. It resulted from the suppressed $|\lambda_{uu}|,|\lambda_{dd}|$
and that $|\lambda_{ss}|$ is larger than
$|\lambda_{uu}|,|\lambda_{dd}|$ two order of magnitudes.

 The functions of $\tau^- \to \mu^- \pi\pi$ decays versus
model parameters are presented in Fig.2, where the left figure is
$Br(\tau^- \to \mu^- \pi^+\pi^-)$ versus $|\lambda_{uu}|$ and
$|\lambda_{dd}|$, and the right is $Br(\tau^- \to \mu^- \pi^0\pi^0)$
versus $|\lambda_{uu}|$ and $|\lambda_{dd}|$. Both decay amplitudes
are relevant to $|\lambda_{uu}|$ and $|\lambda_{dd}|$. When
$|\lambda_{\tau\mu}|=5$, their decay ratios move upward with
$|\lambda_{uu}|$ and $|\lambda_{dd}|$ and both extend the order of
$10^(-14)$. We find that $Br(\tau^- \to \mu^- \pi^+\pi^-)$ climbs
more rapidly than $Br(\tau^-\to \mu^- \pi^0\pi^0)$. It is because
the contributions of $|\lambda_{uu}|$ and $|\lambda_{dd}|$ to the
former amplitude are larger than those to the latter amplitude.

Then, we take $|\lambda_{uu}|=|\lambda_{dd}|=0.5,|\lambda_{ss}|=100$
and analysis the relation of branching ratios versus
$|\lambda_{\tau\mu}|$. Fig.3 gives four decays branching ratios
versus $|\lambda_{\tau\mu}|$ for other fixed parameters. The left
figure is branching ratios for $\tau^- \to \mu^- K \bar{K}$, where
the solid line stands for $\tau^- \to \mu^- K^+K^-$, the dashing
line for $\tau^- \to \mu^- K^0\bar{K}^0$. The experimental data for
$\tau^- \to \mu^- K^+K^-$ and $\tau^- \to \mu^- K^0\bar{K}^0$ are
denoted by the lower horizon line and upper horizon dash line,
respectively. From left figure, we could see the lower horizon line
constraints $|\lambda_{\tau\mu}|$ at $\simeq 4.5$ for $Br(\tau^- \to
\mu^- K^+K^-)$, and the upper limit for $Br(\tau^- \to \mu^-
K^0\bar{K}^0)$ constraints $|\lambda_{\tau\mu}|$ at the order of
$O(1)$. It should be noted that the latest experimental value of
$Br(\tau^- \to \mu^- K^+K^-)$ is based on 671 $fb^{-1}$ data. If the
data of $Br(\tau^- \to \mu^- K^0\bar{K}^0)$ have been updated, the
bound obtained will be more respective. The right figure is $Br$ for
$\tau^- \to \mu^- \pi\pi$ where the solid line stands for $\tau^-
\to \mu^- \pi^+\pi^-$ and the dashing line for $\tau^- \to \mu^-
\pi^0\pi^0$. The experimental data for $\tau^- \to \mu^- \pi^+\pi^-$
and $\tau^- \to \mu^- \pi^0\pi^0$ are denoted by the lower horizon
line and upper horizon dash line, respectively. From the right
figure, we could find $Br(\tau^- \to \mu^- \pi^+\pi^-)$ is more
sensitive to $|\lambda_{\tau\mu}|$ than $Br(\tau^- \to \mu^-
\pi^0\pi^0)$. In all, the $\tau^- \to \mu^- K^+K^-$ decay would make
tighter constraints on the Higgs couplings than those from decays.

\section{\bf Conclusion}
Sum up, we have calculated the branching ratios of $\tau^-
 \to \mu^- PP(PP=K^+K^-,K^0\bar{K}^0,\pi^+\pi^-,\pi^0\pi^0)$
 decays in the model III 2HDM. The neutral Higgs bosons
contribute to theses decays at the tree-level. Comparing to the
$\tau^-\to \mu^- P$ decays, the resonances play a part in $\tau^-\to
\mu^- PP$ processes and to which the massive Higgs bosons have
insensitivity. Only the scalar currents contribute to these decays.
Our work suggests that the parameter $|\lambda_{\tau\mu}|$ is
constrained at the order of $O(1\sim 10^{3})$ by the experimental
data. And in the rational parameters space, their branching ratios
can reach the experimental values. The $\tau^- \to \mu^- K^+K^-$
decay is most sensitive to $|\lambda_{\tau\mu}|$. Our study is hoped
to supply good information for the future experiment and explore the
structure of the 2HDM III model.

%

\begin{acknowledgments}
The work is supported by National Science Foundation under contract
No.10547110, He¡¯nan Educational Committee Innovative Research Team
Foundation under contract No.2010IRTSTHN002, He¡¯nan Educational
Committee Foundation under contract No.2007140007.

\end{acknowledgments}

\end{document}